\documentclass[%
 aip,
 jmp,
 amsmath,amssymb,
 reprint,%
]{revtex4-1}
\usepackage{bbding}
\usepackage{verbatim}
\usepackage{arev}
\usepackage{amssymb}
\usepackage[T1]{fontenc}
\usepackage{array}
\usepackage{graphicx}
\usepackage{dcolumn}
\usepackage{bm}

\usepackage[utf8]{inputenc}
\usepackage[T1]{fontenc}
\usepackage{mathptmx}
\usepackage{etoolbox}

\makeatletter
\def\@email#1#2{%
 \endgroup
 \patchcmd{\titleblock@produce}
  {\frontmatter@RRAPformat}
  {\frontmatter@RRAPformat{\produce@RRAP{*#1\href{mailto:#2}{#2}}}\frontmatter@RRAPformat}
  {}{}
}%
\makeatother
\begin{document}

\preprint{AIP/123-QED}

\title[Atomically-Thin Transition Metal Dichalcogenide Nanolasers: Challenges and Opportunities]{Atomically-Thin Transition Metal Dichalcogenide Nanolasers: Challenges and Opportunities}
\author{Teresa López-Carrasco}%
 \email{m.t.lopez.carrasco@rug.nl}
\author{Marcos H. D. Guimar\~aes}%
 \email{m.h.guimaraes@rug.nl}
\affiliation{ 
Zernike Institute for Advanced Materials, University of Groningen, The Netherlands
}%

\date{\today}

\begin{abstract}
Low energy consumption nanolasers are crucial for advancing on-chip integrated optical interconnects and photonic integrated circuits. 
Monolayer transition metal dichalcogenides (TMDs) have emerged as an energy-efficient alternative to traditional semiconductor materials for nanolaser optical gain medium, promising ultralow lasing threshold powers.
While several studies suggest that TMDs meet the criteria for lasing, whether true lasing has been achieved remains a topic of heavy debate within the scientific community.
In this perspective, we offer an overview of the field, outlining the key characteristics of laser light and methods for testing these properties in TMD-based devices. 
We then conduct a thorough review of recent reports claiming lasing, assessing the findings against established criteria for laser light emission.
Finally, we discuss future research directions and applications, highlighting the key challenges that must be addressed to realize practical TMD-based nanolasers.
\end{abstract}

\maketitle

Lasers have a broad range of applications, including optical interconnects, on-chip light sources, and optical sensors \cite{OpticalInterconnects, Zhou2015, Gauglitz2005}.
When coupled with fiber optics, lasers can transmit information over long distances by sending pulses of infrared or visible light, making them a cornerstone of modern communication technology.
On-chip lasers generate optical signals that are then processed by photonic components such as modulators, detectors, and waveguides, all integrated within the same chip.
This signal encodes data that can be transmitted between different parts of a circuit or across multiple chips, paving the way towards photonic integrated circuit (PICs) technology.   
These systems present numerous advantages over conventional electronic integrated circuits, such as lower power consumption, faster processing speeds, higher interconnect bandwidth, and minimized thermal effects, while maintaining compatibility with existing CMOS fabrication processes \cite{1395890}.

Most PIC components can be fabricated using silicon.
However, silicon has limitations as a light source due to its indirect bandgap, resulting in long-lived free carriers that hinder performance compared to direct bandgap materials \cite{Dutt2024}. 
Additionally, its transparency is confined to the near-infrared region, making it ineffective in the visible spectrum.
Therefore, there is a growing demand for the development of new materials that combine high quantum efficiency, low lasing thresholds, and broad frequency tunability. 

\begin{figure*} [hbt!]
\includegraphics [scale=0.55]{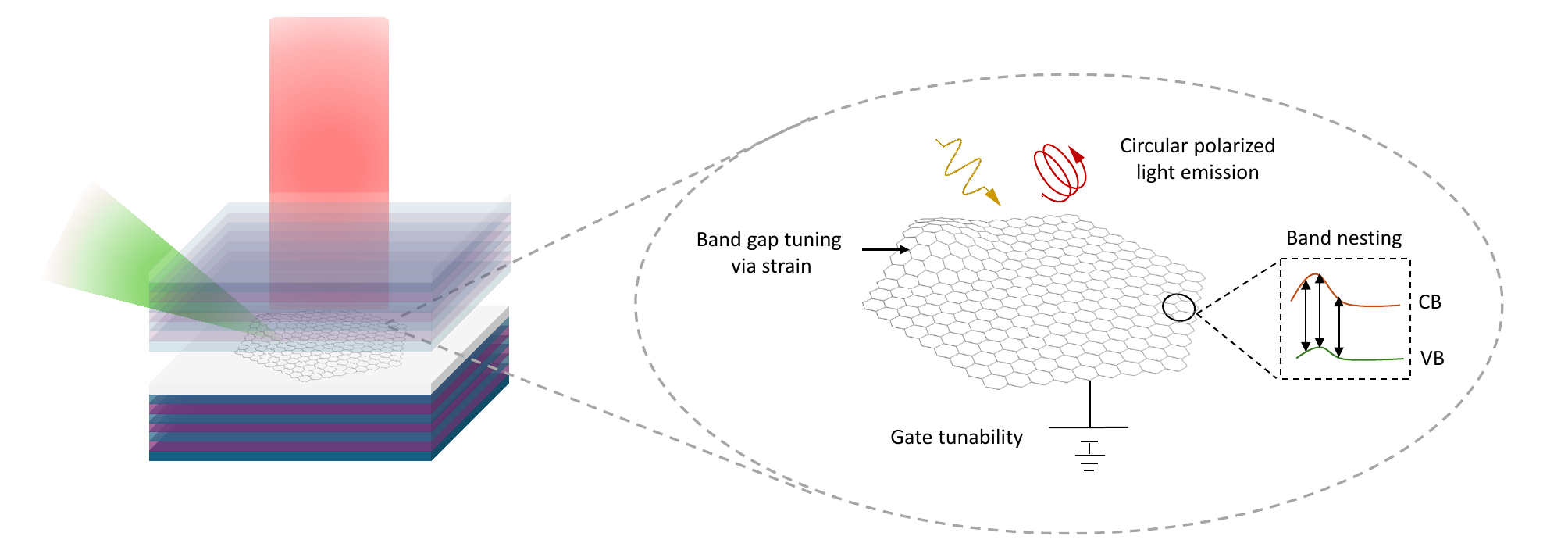}
\caption{\label{fig:fig1} Laser light emission and possible modulation mechanisms of TMD monolayers. Their bandgap and/or their optical absorption can be tuned by electrical gating, strain engineering, and band nesting. TMDs additional valley degree of freedom allows circularly polarized light emission, leading to possible applications on spin/valley lasers.}
\end{figure*}

Traditional materials used in the active regions of nanolasers, such as GaAs, AlGaAs, GaP, InGaP, GaN, InGaAs, InP, and GaInP, tend to show a degradation of their optical properties with reducing dimensionality and do not offer a high tunability of their light emission which can be advantageous for integration.
Transition metal dichalcogenides (TMDs) present a compelling alternative as efficient optical gain media for nanoscale lasers \cite{Salehzadeh2015}. 
One of the key advantages of monolayer TMDs over conventional bulk semiconductors is the strong binding energy of their exciton modes.
Their low dimensionality leads to strong Coulomb interactions between electrons and holes, forming tightly bound excitons, resulting in an intense light-matter interaction \cite{Wilson2021}. 
This interaction can be fine-tuned using methods such as proximity effects, strain, magnetic fields, and electrostatic gating \cite{mann20142, Jauregui2019, Norden2019}, Fig. \ref{fig:fig1}. 
Additionally, TMDs can be combined with other two-dimensional (2D) materials forming heterostructures with highly customizable properties. 
These heterostructures facilitate the development of interlayer exciton-based nanolasers, which exhibit longer exciton lifetimes, reduced non-radiative recombination, and lower lasing thresholds compared to intralayer excitons \cite{Rivera2018}. 
Moreover, TMDs can emit circularly polarized light, enabling novel possibilities for encoding spin information into light polarization -- i.e. spin/valley lasers \cite{https://doi.org/10.1002/advs.202206191}. 
Building on these advantages, monolayer TMDs are one of the most promising platforms for developing tunable light emitters, particularly for optical interconnects and advanced photonic applications.
Moreover, since TMDs have been attracting significant attention from the CMOS industry, with proposals of being included in transistor technology in the late 2030s \cite{Kar2024}, it is natural that we also explore their potential for photonic integrated circuitry.

Although the first reports of TMD-based lasers emerged nearly a decade ago \cite{Ye2015, Wu2015}, there is an ongoing debate about whether these devices can be classified as lasers or if artifacts are merely mimicking lasing behavior. 
This issue was highlighted several years ago in a review by Reeves et al. \cite{Reeves2018}, and Nature Photonics has emphasized the importance of rigorously verifying the properties of laser light in such claims \cite{Reportingchecklistformanuscriptswithaclaimoflasing2017}. 
In recent years, many advances have been made in the field, including reports of TMD-based nanolaser devices that meet all the established criteria for laser light emission, including circularly polarized nanolasers\cite{Barth, Rong2023}. 
These developments underscore the high potential of TMD-based lasers for future atomically-thin optoelectronic applications.

In this perspective, we provide a broad overview of the field, highlighting the key properties of laser light and the methods for evaluating them in devices using TMD as the gain medium. 
We also present a critical analysis of previous studies claiming lasing behavior in TMD devices, applying the established laser light criteria to critically assess their results. 
Finally, we outline future research directions and potential applications of TMD-based nanolasers, emphasizing the major challenges that must be addressed to bring these technologies to fruition.

\begin{figure*} [hbt!]
\includegraphics [scale=0.89]{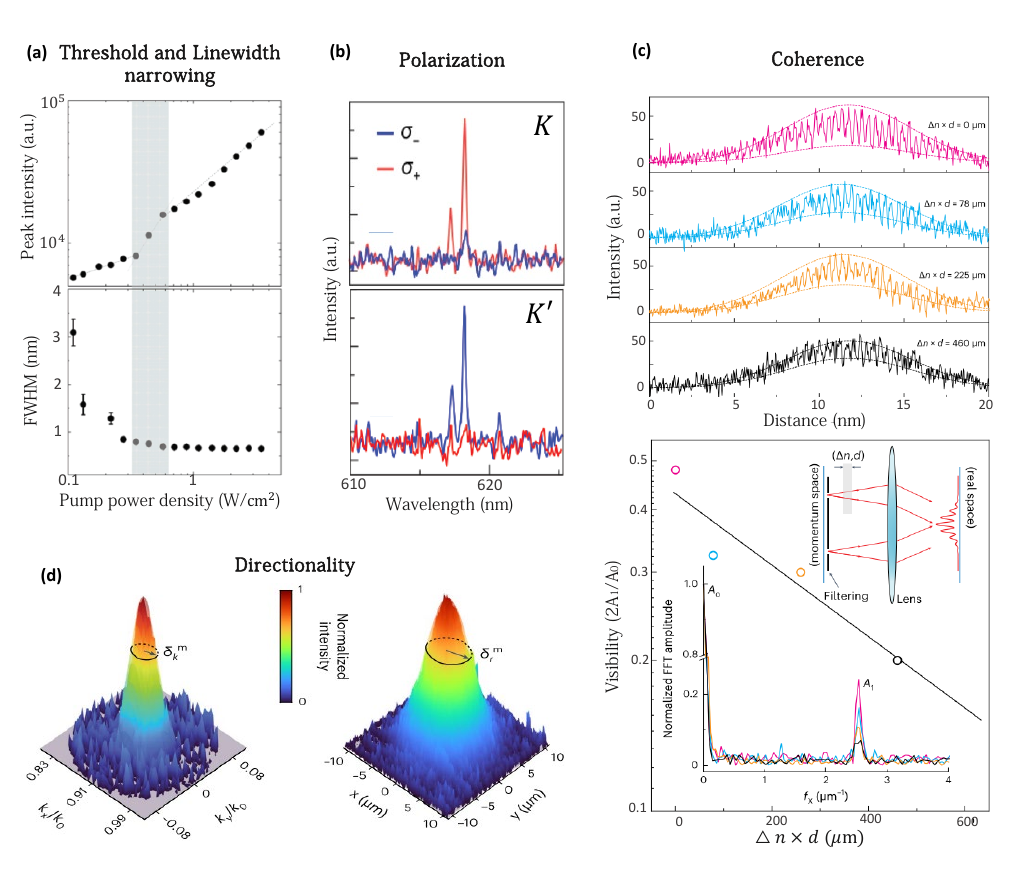}
\caption{\label{fig:lasing_cirteria} Assessment Methods for Lasing Criteria. 
(a) Output photoluminescence peak intensity and full width at half-maximum (FWHM) for different pump powers showing a clear "kink" for the intensity and beam at the threshold, indicated by the shaded gray region. Figure adapted from \cite{Barth}.
(b) Circular polarization modulation of the light emission for optical pumping at different spin-valley resonances into a heterostructure microcavity. Figure adapted from \cite{Rong2023}. 
(c) Top: Two-beam interference fringes with varying time delays ($\Delta n \times d$). Dashed curves represent Gaussian fits to the fringes' envelopes. Bottom: Visibility values with a linear fit to the decaying visibility. Top inset: Interference setup schematic. Bottom inset: FFT amplitudes for different positive spatial frequencies. Figure adapted from \cite{Rong2023}.
(d) Average spatial profiles along $k_{x}$ and $k_{y}$ of the far-field emission patterns to determine the beam divergence angle. Figure adapted from \cite{Rong2023}.}
\end{figure*}

For a light source to be considered a laser it needs to fulfill specific characteristics \cite{Wright1980}.
Here we adopt the comprehensive checklist proposed by Nature Photonics in 2017 \cite{Reportingchecklistformanuscriptswithaclaimoflasing2017} to scrutinize the literature on TMD-based lasers.
This checklist covers all the main laser light characteristics and provides a critical view of the field.
The characteristics of a laser are: (i) the presence of a threshold behavior, (ii) linewidth narrowing above the threshold, (iii) highly polarized light emission, (iv) highly coherent light, and (v) directional light emission.
Fig.~\ref{fig:lasing_cirteria} illustrates the characterization of all these criteria for different studies from the literature. 
Additionally, we point out that it is important to estimate the $\beta$-factor that quantifies the importance of other non-coherent light-emitting processes, e.g. amplified spontaneous emission  \cite{Lippi2021}.
We summarize the characteristics above for several works in the literature in Table \ref{tab:lasing_charact} and their device specifications in Table \ref{tab:nanoresonators_charact}.
Below we discuss each of the lasing criteria and give examples on how to characterize them.

The most prominent evidence of laser light is the presence of a threshold behavior on the light emission intensity versus pumping power.
Note that all lasers based on TMDs reported so far use optical pumping.
This is mostly due to the difficulty in producing low-resistance ohmic contacts to TMD monolayers \cite{Chhowalla2021}.
The threshold behavior takes place when the gain of the optical mode is equal to the losses of the cavity \cite{Ning2013}.
This behavior is considered the first evidence of lasing in several works \cite{Ye2015, Barth, Fang2019}.
The identification of the threshold of optically pumped lasers can be performed using the so-called ($L-L$) plots, where light in \textit{vs} light out is plotted, Fig.~\ref{fig:lasing_cirteria}(a). 
These plots are expected to show a clear change in the slope of the outgoing light intensity with increasing pumping power.
This so-called "kink" corresponds to a transition from amplified spontaneous emission to stimulated emission.
Such a test is present in all TMD laser reports as the main proof for the laser behavior of their devices (Table~\ref{tab:lasing_charact}).
However, for nanolasers this criterion in itself is not sufficient, since the contribution of spontaneous emission to the cavity mode is not negligible \cite{Lippi2021}.
Particularly, in single-mode cavity systems there is a significant increase in the amplified spontaneous emission, which could also show a threshold behavior \cite{app9214591}.
Therefore, the observation of a kink in the $L-L$ plot cannot be used as the sole evidence of a transition to amplified stimulated emission.
Bjork et al.\cite{Mg1994} proposed a more precise criterion for identifying the lasing threshold in nanolasers, known as the Quantum Threshold (QT) condition.
This criterion states that the threshold is reached when stimulated emission surpasses spontaneous emission. 
The QT condition corresponds to a situation in which the average photon number in the mode is unity.
From this, the power that the laser should emit -- i.e. the output power -- at threshold ($P_{out}$) can be estimated.
To obtain this value, we divide the energy of the photon ($E = \frac{hc}{\lambda}$) by its average lifetime in the optical resonator ($\tau$), thus maintaining the one-photon average inside the cavity.
Here $\lambda$ is the emitted photon wavelength, $c$ the speed of light, and $h$ the Planck's constant.
We can determine $\tau$ and use it to estimate $P_{out}$ as follows:

\begin{equation}
    \begin{split}
    \tau=\frac{Q T_{cycle}}{2\pi}, \\
    P_{out}= \frac{E}{\tau} = h2\pi c^{2}/(\lambda^{2}Q). 
\end{split}
\end{equation}

\noindent where $Q$ is the quality factor of the optical resonator and $T_{cycle} = \lambda/c$ is the period of the photon in the resonator.
As the spot size of the pump laser is not always reported in the literature, we assume, when necessary, a diffraction-limited spot, obtained from the given numerical aperture (NA) of the focusing lens/objective, to evaluate the incident pumping power ($P_{in}$).
The computed output power is then compared to the input threshold power reported by the authors.
In some publications, our calculations give $P_{out}>P_{in}$, which violates the conservation of energy.
This means that, following the QT condition, the observation does not correspond to a lasing threshold \cite{Qian2024, Yu2022, Sung2022, Paik2019, Liu2019, Zhao2018, Shang2017, Li2017, Wu2015}. 
On the other hand, the kink in the $L-L$ plot could correspond to the threshold if $P_{out}<P_{in}$. 
This condition can however not be verified when the output power is not reported (as in Refs. \cite{Lin2023, Ge2019, Liao2019, Fang2018,Salehzadeh2015, Ye2015, Koulas-Simos2024, Rong2023}). 
So far, two studies have reported experimental evidence with measurements of $P_{out}$ \cite{Barth, Rong2023}. 
One major challenge has been achieving sufficiently high output power in these devices to be experimentally measured. 
Researchers are addressing this issue through various approaches, such as employing dual-resonant plasmonic structures \cite{Barth}.

To achieve ultra-low threshold systems, the ratio of the energy stored in the nanoresonator to the energy lost in a photon cycle should be as high as possible. 
This is quantified by the quality factor: 
\begin{equation}
    Q=\frac{\omega_{m}}{FWHM}, 
\end{equation}
where $\omega_{m}$ is the resonant frequency and FWHM the full width at half maximum of the mode.
Table~\ref{tab:nanoresonators_charact} gives an overview of the nanoresonators used in different works.
We have classified them into three main groups: whispering galleries, photonic crystal cavities and vertical cavity surface-emitting lasers (VCELS).

The second key condition used to assess the emission of laser light is the linewidth narrowing of the output beam, which by convention should be reduced by a factor of two when the threshold is achieved \cite{Reeves2018}. 
A perfect laser emits only one specific wavelength, however real lasers produce light with a certain bandwidth.
This deviation is usually measured by the FWHM of the emitted light spectrum peak. 
At the transition from spontaneous to stimulated emission, the FWHM decreases sharply, which is called linewidth narrowing.
Another factor that influences the FWHM is the $Q$-factor of the nanoresonator. 
The higher the $Q$-factor, the narrower the spectrum of the optical mode and therefore the smaller the output beam's spectral width.
Furthermore, this linewidth narrowing is correlated to the temporal coherence. 
A smaller peak width indicates a higher coherence time.
The way to show this peak width reduction is by taking the output beam spectrum at different laser pump powers (Fig.~\ref{fig:lasing_cirteria}(a)). 
The drop in the FWHM is an indication of the lasing threshold.
This study is performed in all the TMD laser reports, nevertheless, only a few of them fulfill criterion of a factor of two reduction \cite{Barth,Ge2019, Liao2019, Shang2017, Li2017, Salehzadeh2015, Rong2023}.

One of the consequences of a stimulated emitted light source is the high degree of polarization of the emitted photons.
Therefore, the polarization of the output beam is another parameter used to characterize lasing. 
Eight of the nineteen works on TMD lasers report the polarization of their laser beam \cite{Barth,Qian2024,Lin2023,Sung2022,Paik2019,Fang2018,Li2017,Salehzadeh2015,Wu2015, Rong2023} and only one explores the circular polarization of the emitted light \cite{Rong2023}.
Fig.~\ref{fig:lasing_cirteria}(b) shows an example of a circular polarized laser beam, right or left handed depending on the valley from where the photons are generated.
The circular polarization of the output beam is significantly influenced by the intervalley scattering of excitons, with the scattering time ($\tau_{iv}$ ) typically on the order of picoseconds. 
For the beam to achieve circular polarization, all photons must result from exciton recombination within the same valley.
Therefore, if the photon lifetime in the cavity is longer than $\tau_{iv}$, photons will not have sufficient time to interact and create population inversion before electrons scatter to a different valley.
To effectively modulate the circular polarization of the output beam, $\tau_{iv}$ needs to be at least in the picosecond range.
Consequently, achieving photon lifetimes longer than $\tau_{iv}$ requires quality factors beyond 1300.

\begin{table*} [t!]
\caption{\label{tab:lasing_charact} Verification of the lasing conditions for all the TMDs-based spin lasers reported so far in the literature.
A $\ballotcheck$ ($\ballotx$) means that the devices do (not) fulfill the condition.
A $-$ indicates that the criterion is not reported by the authors.
In addition, for the quantum threshold condition, $(\ballotcheck \ballotcheck)$ means that $P_{in}>P_{out}$ for the calculated and the reported $P_{out}$, only one $(\ballotcheck)$ means $P_{in}>P_{out}$ but the authors do not report the $P_{out}$ value.
Finally, $(\ballotx)$  means $P_{in}<P_{out}$  so they do not fulfill the quantum threshold (QT) condition.}
\begin{ruledtabular}
\begin{tabular}{ccccccccc}

 Gain medium & T (K) & Threshold (QT)& Linewidth &Coherence & Polarization & Directionality & Year& Ref.\\
 &  & W/cm$^{2}$&  narrowing (fact. 2)& &  &  & & \\
\hline 
WSe$_{2}$ (1L) & RT &10200 $(\ballotcheck)$ & $\ballotcheck$ $(\ballotx)$&$\ballotcheck$ & $-$& $-$&2024 &\cite{Koulas-Simos2024} \\
WS$_{2}$ (1L) & RT&  $<1 (\ballotcheck\ballotcheck)$ & $\ballotcheck$ $(\ballotcheck)$ & $\ballotcheck$  &$\ballotcheck$ & $\ballotcheck$  &2024& \cite{Barth}\\
MoSe$_{2}/$WSe$_{2}$ (2L) & 10  & $\sim13$ $(\ballotx)$ & $\ballotcheck$ $(\ballotx)$ & $\ballotcheck$ & $\ballotcheck$ & $-$ & 2024 &\cite{Qian2024}  \\
WS$_{2}$ (1L) & RT & 2000 $(\ballotcheck \ballotcheck)$& $\ballotcheck$ $(\ballotcheck)$ & $\ballotcheck$  &$\ballotcheck$ & $\ballotcheck$  &2023& \cite{Rong2023} \\
MoS$_{2}/$WSe$_{2}$ (2L)& RT & $\sim 5000$ $(\ballotcheck)$  & $\ballotcheck$ $(\ballotx)$ & $\ballotcheck$ \footnote{Correlation measurements are not performed because of the laser low signal. Moiré excitons are used.} & $\ballotcheck$ & $-$ & 2023&\cite{Lin2023} \\
WSe$_{2}$ (2L) & RT & $0.72$ $(\ballotx)$ & $\ballotcheck (\ballotx)$& $\ballotcheck$ & $-$ & $-$ & 2022 &\cite{Yu2022} \\
WS$_{2}$ (50 nm)& RT & $1250$ $(\ballotx)$ & $\ballotcheck$ $(\ballotx)$ &$\ballotcheck$ &$\ballotcheck$& $-$ & 2022 &\cite{Sung2022} \\
MoTe$_{2}$ (1L) & RT & 4200  $(\ballotcheck)$&$\ballotcheck$ $(\ballotx)$& $-$ &$-$ &$-$ & 2019 &\cite{Fang2019} \\
WS$_{2}$ (1L)& RT & 116$(\ballotx)$ &$\ballotcheck$ $(\ballotcheck)$ & $-$ & $-$ & $\ballotcheck$ & 2019 &\cite{Ge2019}\\
WSe$_{2}/$MoSe$_{2}$ (2L) &70  & $0.18\:\mu $W $(\ballotx)$& $\ballotcheck$ $(\ballotx)$& $\ballotcheck$ \footnote{Only spatial coherence is shown.} &$\ballotcheck$ & $-$ & 2019 &\cite{Paik2019}\\ 
MoS$_{2}$ (1L) & RT & $70$ $(\ballotcheck)$ & $\ballotcheck$ $(\ballotcheck)$ & $-$ & $-$ &  $-$ &2019 &\cite{Liao2019}\\
MoS$_{2}/$WSe$_{2}$ (2L) & RT & $\sim688$ $(\ballotx)$ & $\ballotcheck$ $(\ballotx)$ & $-$ \footnote{The coherence is demonstrated at 5 K.} &$-$ &$-$ & 2019&\cite{Liu2019}\\
MoS$_{2}$ (1L) & RT & $380 (\ballotx)$ \footnote{Lasing is demonstrated for a temperature of 77 K with a threshold of 32 W/cm$^{2}$ and at a temperature of 400 K with a threshold of 580 W/cm$^{2}$.} &$\ballotcheck$ $(\ballotx)$ & $-$ & $-$ & $-$ & 2018&\cite{Zhao2018}  \\
MoTe$_{2}$ (220 nm) & RT& $1500(\ballotcheck)$ & $\ballotcheck$ $(\ballotx)$ \footnote{Linewidth narrowing of 0.02 nm. The output power of the laser is measured and has a value of $\sim20$ pW/nm.}& $-$ & $\ballotcheck$ & $-$ & 2018 &\cite{Fang2018}\\
WS$_{2}$ (1L)& RT & $0.44$ $(\ballotx)$& $\ballotcheck$ $(\ballotcheck)$ & $\ballotcheck$ &$-$&$\ballotcheck$ & 2017&\cite{Shang2017}\\ 
MoTe$_{2}$ (1L) & RT&  $6.6$ $(\ballotx)$ & $\ballotcheck$ $(\ballotcheck)$& $-$& $\ballotcheck$&$-$ & 2017 &\cite{Li2017}\\
MoS$_{2}$ (1L)& RT & $\sim5 \mu$W $(\ballotcheck)$  & $\ballotcheck$ $(\ballotcheck)$ &$-$ &$\ballotcheck$ & $-$ &2015 &\cite{Salehzadeh2015}\\
WSe$_{2}$ (1L)& 130& $\sim1$ $(\ballotx)$ & $\ballotcheck$ $(\ballotx)$ & $-$ & $\ballotcheck$ & $-$ &2015 &\cite{Wu2015}\\
WS$_{2}$ (1L) &  10 & $22400000$ $(\ballotcheck)$ &$\ballotcheck$$(\ballotx)$ &$-$ &$-$ &$-$ & 2015&\cite{Ye2015}  \\

\end{tabular}
\end{ruledtabular}
\end{table*}

Lasing criteria indicate that the emitted light must be highly coherent in both space and time domains. 
In the low carrier injection regime only spontaneous emission occurs, which means that the light is incoherent and non-directional. 
Coherence arises when the higher injection regime is reached, leading to population inversion and resulting in stimulated emission. 
We analyze the spatial and temporal coherence of the light emission. 
Spatial coherence describes the correlation between waves at different points in space, either laterally, or longitudinally while temporal coherence details the correlation between waves observed at different times.
Some authors use Young's double slit interference experiment or the Michelson interferometer to characterize the spatial coherence of the output beam \cite{Barth,Rong2023}.    
The resolution of the two-slit interference pattern is directly related to the degree of coherence of the waves. 
A large source that is not collimated or that mixes many different wavelengths will have lower contrast for the interference pattern, denoted as "visibility". 
Fig.~\ref{fig:lasing_cirteria}(c) shows how the visibility increases as the system goes beyond threshold.
Nine works experimentally verified the coherence of their laser beams \cite{Barth,Qian2024,Lin2023,Yu2022,Sung2022, Paik2019, Shang2017, Rong2023, Koulas-Simos2024}.


The last criterion we consider is the directionality of the emitted light.
One method to assess the beam’s directionality is through imaging the back focal plane of the lens to measure the wave vector distribution.
This technique is known as Fourier plane imaging or back focal plane imaging (BFP) \cite{Vasista_2019}.
In Fig.~\ref{fig:lasing_cirteria}(d), we show an example of the far-field emission pattern of the output beam \cite{Rong2023}. 
By fitting the profile along k$_{x}$ with a Gaussian and measuring its width, the scattering behavior of the laser beam can be quantified.
Introducing a gain medium within a nanocavity confines the direction of transmitted photons, thereby enhancing the beam's directionality. 
Although diffraction will still cause some divergence, even in a fully spatially coherent beam, diffraction is the sole factor limiting the laser’s directionality, under optimal operating conditions.
We are aware of four studies on TMD lasers that have reported on the directionality of the output beams of their devices \cite{Barth, Ge2019, Shang2017, Rong2023}. 
This is largely due to the low output power, which limits the ability to perform beam profile through BFP measurements.
\begin{figure} [t!]
    \centering
    \includegraphics[width=0.80\linewidth]{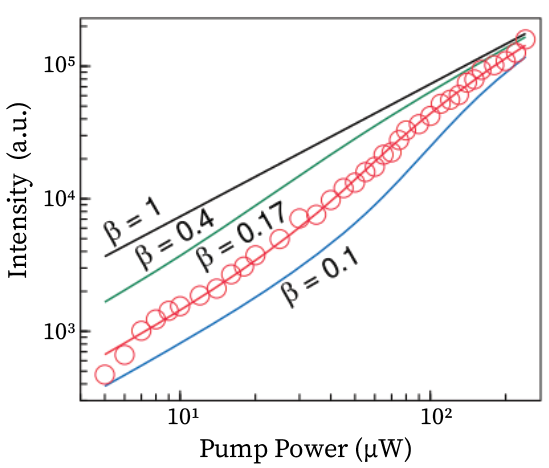}
    \caption{$\beta$-factor estimation. Output beam intensity as a function of pump power with fits for different beta values. Figure adapted from Ref. \cite{Liu2019}.}
    \label{fig:bETA}
\end{figure}
An additional factor to consider when characterizing a laser beam is the number of photons generated by spontaneous emission that couple with the cavity mode.
Although this is not strictly a lasing criterion, it gives additional information on the importance of amplified spontaneous emission in $L-L$ plots.
This factor is quantified by the so-called $\beta$-factor, which indicates the proportion of output power derived from amplified stimulated emission.
Therefore, a $\beta$-factor = 1 represents that all photons generated by spontaneous emission couple to the cavity mode.
In practice, the $\beta$-factor is often determined by fitting experimental data to various laser rate equations\cite{Yokoyama1989}.
An example, obtained in Ref. \cite{Liu2019}, is shown in Fig.~\ref{fig:bETA}. 
A high $\beta$-factor can enhance the device's output power, effectively reducing the laser threshold.
However, these incoherent photons also compromise the coherence of the output laser beam, leading to a debate on whether a high $\beta$-factor is desirable, since it is highly dependent on the specific application of the device.

Table~\ref{tab:lasing_charact} reveals that all reported TMD lasers to date exhibit threshold behavior and linewidth narrowing.
According to our estimations, the QT condition is not fulfilled by most of these studies, nor is the expected factor of two reduction in emission linewidth.
From the reported data, it is evident that the most challenging properties to measure are coherence, polarization, and directionality. 
In some cases, these cannot be fully assessed due to the low output power of the devices.
However, we point out that two works fulfill all the criteria for laser light emission \cite{Barth,Rong2023}.
These works set the stage for the development of CMOS-compatible ultralow-threshold TMD lasers.
In the following sections, we give a perspective on the possible future directions of the field.
In conventional nanolasers, a light source (pump) excites electrons from the valence to the conduction band.
These excited carriers then undergo non-radiative decay until they reach the lasing energy level, creating a population inversion that leads to stimulated emission.
Importantly, TMDs exhibit a unique electronic structure where the conduction band has two minima (valleys) of equal energy but located at opposite positions in momentum space.
These valleys can be individually addressed by circularly polarized light with opposite helicities.
This selective valley excitation introduces a new degree-of-freedom -- the valley index --which forms the foundation of an emerging field known as valleytronics \cite{valleytronics}.
Additionally, since the excited carriers can potentially accumulate within a single valley, the number of carriers required to achieve population inversion is significantly reduced, thereby potentially lowering the lasing threshold \cite{Zeng2012}.

Based on this particular property of TMDs, the work by Rong et al.\cite{Rong2023} reported a valley-polarized laser operating at room temperature using monolayer WS$_{2}$. 
This laser, which emits at a wavelength of 618.2 nm, achieved a lasing threshold of 2000 W/cm$^{2}$.
The authors observe that by incorporating a WS$_{2}$ monolayer on a heterostructure microcavity, designed to support spin-valley resonance, they can obtain spin-dependent emission (Fig.~\ref{fig:lasing_cirteria}(b)).
Their findings demonstrate the potential for a laser in which the valley where carriers are excited can be selectively chosen, emitting light with opposite circular polarizations.  

Due to the high spin-orbit coupling and lack of inversion symmetry in monolayer TMDs, their valley degree-of-freedom is strongly linked to the spin direction of the carriers.
This property allows TMDs to tackle one of the key challenges in spintronics, the efficient transport of spin information over long distances. 
Spin transport has been studied in a variety of materials, including metals, semiconductors, and graphene, primarily through the diffusion of conduction electrons \cite{Fabian2007}, and in magnetic insulators via the diffusion of non-equilibrium magnons \cite{Cornelissen2015}. 
However, these methods typically achieve spin transport distances on the order of micrometers.
This is where TMD-based valley/spin lasers come into play, offering the potential to encode spin information into light polarization, enabling the transport of spin information over much longer distances \cite{ZUTIC2020113949,Lindemann2019}.
Additionally, spin lasers offer much higher polarization modulation frequencies when compared to conventional lasers.
Modulation frequencies up to 100's of GHz have been demonstrated for conventional VCSEL spin lasers \cite{Lindemann2019}, opening an avenue for high bandwidth data transmission.
Such high modulation bandwidth is made possible through an anisotropy on the index of refraction of the material along different crystal directions.
This was obtained by mechanically bending the VCSELs.
We envision that spin lasers based on 2D materials could take the advantage van der Waals heterostructure engineering to obtain the anisotropy in the index of refraction by design, without the need of mechanical means.

\begin{table*} [hbt!]
\caption{\label{tab:nanoresonators_charact} Type of nanoresonators used in the papers reported in table~\ref{tab:lasing_charact}, and their main characteristics.}
\begin{ruledtabular}
\begin{tabular}{cccc}

Microcavity & Quality factor  & Comments& Ref. \\ \hline
VCSEL&$\sim230$&& \cite{Koulas-Simos2024} \\
Dual resonance dielectric metasurface & $\sim1700 - 3000$ & \footnote{CVD-grown monolayer. Spatial coherence of 30 um is shown.} & \cite{Barth}\\
Photonic crystal cavity&$\sim12500$ &  &\cite{Qian2024} \\
Photonic crystal cavity & $\sim5600$& & \cite{Rong2023}\\
Silicon topological nanocavity & $\sim 10000$  &   & \cite{Lin2023}  \\
Whispering gallery  & $\sim200$  &   &\cite{Yu2022} \\
Whispering gallery & $\sim400$&  &\cite{Sung2022}\\
Photonic crystal cavity & $\sim4500$ &  & \cite{Fang2019}\\
Photonic crystal lattice  &$\sim2500$ &  &\cite{Ge2019} \\
Photonic crystal cavity   & $\sim\:630$ &  \footnote{The TM-polarized emission is not coupled to the cavity mode. } &\cite{Paik2019}\\
Whispering gallery & $\sim350$&  &\cite{Liao2019}\\
Photonic crystal cavity & $\sim 423$&  &\cite{Liu2019} \\
Whispering gallery & $\sim740$  & & \cite{Zhao2018} \\
Photonic crystal nanocavity & $\sim5900$ &  & \cite{Fang2018}\\
VCSEL & $\sim1300$ & & \cite{Shang2017} \\ 
Photonic crystal cavity  & $\sim5603$ &  & \cite{Li2017}\\
Whispering gallery & $\sim200 - 400$& \footnote{Laser spot size not specify.} & \cite{Salehzadeh2015} \\
Photonic crystal cavity  & $\sim 2500$   &  \footnote{First report of a 2D material-based nanoscale laser. Linewidth measurements taken at 160 K.}&\cite{Wu2015}\\
Whispering gallery  & $\sim2604$ &\footnote{Two different modes are reported.} &\cite{Ye2015}\\

\end{tabular}
\end{ruledtabular}
\end{table*}

Integrating nanolasers into on-chip optical interconnects requires meeting several criteria simultaneously, including sufficient laser output power, low power consumption, continuous-wave operation, powered electrically at room temperature, highly efficient coupling of the laser output into waveguides, and high modulation rates.
A persistent challenge in TMD-based lasers is their low output power, which complicates the characterization of the laser beam and raises doubts about the true lasing nature of the emitted light. 
To address this, two primary approaches have been explored: enhancing the emission or the absorption of TMD monolayers.

The most straightforward method to boost the photoluminescence (PL) emission of TMD monolayers is to increase the pump power. 
However, this approach faces two major issues: potential damage to the material at high power levels and the enhancement of many-body interactions, such as non-radiative exciton-exciton annihilation, caused by the high carrier density in these atomically thin layers \cite{Yuan2015}. 
Alternative strategies to enhance PL emission include molecular adsorption \cite{Tongay2013, Mouri2013}, defect engineering \cite{Tongay20132, Lee2018}, and strain modulation \cite{Roldan2015}.
Liao et al. \cite{Liao2019} introduced a photoactivation technique to improve the room-temperature quantum yield of monolayer WS$_{2}$ by using optical silica-based micro/nanofibers as resonators. 
The laser heating of these fibers frees oxygen atoms that can fill sulfur vacancies, forming bridges with neighboring atoms, which significantly enhances PL emission. 
This process reduced the lasing threshold from 70 Wcm$^{-2}$ to 5 Wcm$^{-2}$ in monolayer $WS_{2}$. 
However, this study did not report on the output power, coherence, polarization, or directionality of the laser.

Another approach to improving TMD laser output focuses on increasing the monolayer’s absorption.
This could be done by including the use of plasmonic structures, patterned metasurfaces, and increasing the thickness of the 2D layers.
Barth et al. \cite{Barth} created a room-temperature excitonic laser using a monolayer of WS$_{2}$ on a dual-resonance metasurface, enhancing both the absorption and emission wavelengths of the WS$_{2}$ monolayer. 
The rectangular lattice in their design effectively redirects photons horizontally, increasing light-matter interaction and reducing the lasing threshold to an ultra-low 1 W/cm$^{2}$.
This study demonstrates that improving the absorption and emission of the gain medium is a highly effective approach to enhancing laser output power, resulting in one of the lowest lasing thresholds ever reported in nanolasers.
Additionally, a particularly promising method for enhancing absorption is through band nesting, a phenomenon where conduction and valence bands are approximately equispaced over regions in the Brillouin zone. 
In 2D materials, band nesting leads to singularities in the joint density of states, resulting in a strongly enhanced optical response at resonant frequencies \cite{Mennel2020}.
Lee et al.\cite{Lee2023} reported a remarkable 95\% absorption efficiency in TMDs through band nesting at room temperature which could be leveraged for lasing applications.
This could further be combined with optical cavity engineering to bring down the lasing threshold to significantly lower values.

Another promising avenue for the development of TMD lasers is the exploration of van der Waals (vdW) heterostructures through the generation of interlayer excitons. 
One of the key challenges in this approach lies in the fabrication process. 
Precise alignment of the monolayers is essential, as interlayer excitons require the TMDs to be aligned at either 0 or 60 degrees for the matching of their Brillouin zone. 
Any misalignment higher than a few degrees results in a significant reduction in PL intensity \cite{Nayak2017ProbingEO}. 
To overcome this limitation, investigating interlayer exciton emission using $\Gamma$ point transitions offers a promising alternative for future TMD-based lasers \cite{Gatti2023}, as the monolayer orientation would no longer be a critical factor.

Fully harnessing the potential of 2D TMDs requires scalable and reliable synthesis of these atomically thin materials and their heterostructures. 
Significant progress has been made in the scalability of TMD systems \cite{Li2024}. 
While mechanical exfoliation can yield high-quality TMD monolayers with strong PL, its small-scale output and labor-intensive process make it unsuitable for large-scale production. 
Efforts to achieve wafer-scale TMD films have been successful, such as the work of Kang et al. \cite{Kang2015}, who produced  4-inch wafer-scale MoS$_{2}$ films on SiO$_{2}$ wafers with quality comparable to exfoliated samples.
However, the scalable integration of wafer-scale TMDs with optical cavities and photonic circuits without degrading the optical properties of the TMDs is highly challenging.
Therefore, the integration of TMD lasers using, for example, VCSEL geometries would require damage-free deposition of materials on top of TMDs or the improvement of sacalable TMD layer transfer.

Another major challenge for TMD-based lasers is the demonstration of electrical pumping, a requirement to be effectively integrated into photonic integrated circuits. 
This presents a significant challenge due to the difficulty in forming high-quality low-resistance ohmic electrical contacts with atomically-thin semiconductors. 
Wang and Chhowalla et al. addressed this issue in a recent review \cite{Chhowalla2021}, pointing out that while doping is commonly used to improve contact quality in bulk materials, it can destabilize atomically-thin semiconductors and introduce defects.
A viable route is provided by, for example, lateral contacts made of the metallic crystal phase of the 2D semiconductors \cite{Kappera2014, Cho2015}.
Such approach has been shown to significantly improve the opto-electronic properties of vdW materials-based devices\cite{Hidding2024}.

The development of PICs based on TMD lasers and other 2D materials presents an exciting frontier with immense potential for transforming the landscape of optoelectronics. 
Recent advancements in the field, such as improved methods for scalable production of high-quality TMD monolayers, breakthroughs in vdW heterostructure fabrication and engineering, and the successful demonstration of low-threshold nanolasers, have laid a strong foundation for future innovation.
These achievements offer a clear path toward the realization of PICs with significantly lower energy consumption compared to conventional electronic circuits.
Nonetheless, for further developments in the field we point out that claims of TMD lasers need to undergo strong scrutiny with respect to the well-established lasing criteria.
These criteria should be checked by the authors and properly reported, so the field can build on solid basis.
The unique optical properties of TMDs, including their tunable band gaps, strong spin-orbit coupling, and enhanced light-matter interactions, provide the potential for entirely new functionalities that cannot be achieved with traditional semiconductor materials.
 
\begin{acknowledgments}
We would like to thank Daniel Vaquero-Monte and Maxen Cosset-Chéneau for their helpful feedback on this manuscript, and Hussain Alsalman for insightful discussions on band nesting effects in TMDs.
We acknowledge the financial support by the Zernike Institute for Advanced Materials and the European Union (ERC, 2D-OPTOSPIN, 101076932). Views and opinions expressed are however those of the author(s) only and do not necessarily reflect those of the European Union or the European Research Council. Neither the European Union nor the granting authority can be held responsible for them.
\end{acknowledgments}

\end{document}